Tunnel magnetoresistance and spin-transfer-torque switching in polycrystalline $Co_2FeAl$ full-Heusler alloy magnetic tunnel junctions on $Si/SiO_2$ amorphous substrates

Zhenchao Wen, Hiroaki Sukegawa, Shinya Kasai, Koichiro Inomata, and Seiji Mitani

*National Institute for Materials Science (NIMS), 1-2-1 Sengen, Tsukuba 305-0047, Japan*

Abstract:

We studied polycrystalline *B*2-type $Co_2FeAl$ (CFA) full-Heusler alloy based magnetic tunnel junctions (MTJs) fabricated on a $Si/SiO_2$ amorphous substrate. Polycrystalline CFA films with a (001) orientation, a high *B*2 ordering, and a flat surface were achieved using a MgO buffer layer. A tunnel magnetoresistance (TMR) ratio up to 175% was obtained for an MTJ with a CFA/MgO/CoFe structure on a 7.5-nm-thick MgO buffer. Spin-transfer torque induced magnetization switching was achieved in the MTJs with a 2-nm-thick polycrystalline CFA film as a switching layer. Using a thermal activation model, the intrinsic critical current density ($J_{c0}$) was determined to be $8.2 \times 10^6$ A/cm$^2$, which is lower than $2.9 \times 10^7$ A/cm$^2$, the value for epitaxial CFA-MTJs [*Appl. Phys. Lett.* **100**, 182403 (2012)]. We found that the Gilbert damping constant ($\alpha$) evaluated using ferromagnetic resonance measurements for the polycrystalline CFA film was ~0.015 and was almost independent of the CFA thickness (2~18 nm). The low $J_{c0}$ for the polycrystalline MTJ was mainly attributed to the low $\alpha$ of the CFA layer compared with the value in the epitaxial one (~0.04).

# I. INTRODUCTION

Half-metallic ferromagnets (HMFs) draw great interest because of the perfect spin polarization of conduction electrons at the Fermi level, which is considered to enhance the spin-dependent transport efficiency of high-performance spintronic devices. [1–3] Cobalt-based full-Heusler alloys with the chemical formula Co$_2$YZ (where Y is a transition metal and Z is a main group element), are extensively studied as a type of HMFs owing to their high Curie temperature of approximately 1000 K, high spin polarization, and low damping constant. [4, 5] They exhibit great potential for applications in spintronics, including current-perpendicular-to-plane giant magnetoresistance (CPP-GMR) read heads, [6, 7] magnetoresistive random access memories (MRAMs), [8] and spin transistors such as spin-functional metal-oxide-semiconductor field-effect transistors (spin-MOSFETs). [9, 10] In particular, magnetic tunnel junctions (MTJs) with Co-based full-Heusler alloy electrodes have been shown tremendously increasing tunnel magnetoresistance (TMR) ratios during the last decade since Inomata *et al.* [11] demonstrated a TMR ratio of 16% using Co$_2$Cr$_{0.6}$Fe$_{0.4}$Al/AlO$_x$/CoFe MTJs at room temperature (RT). [12–21] Recently, a remarkable TMR ratio of approximately 2000% at 4.2 K (354% at RT) was achieved using epitaxial Co$_2$MnSi/MgO/Co$_2$MnSi(001) MTJs, demonstrating the half-metallicity of Co-based Heusler alloys and strong $\Delta_1$ coherent tunneling effect in the MgO/Heusler MTJs [21].

The Co$_2$FeAl (CFA) alloy is of particular interest because of its high spin polarization (a half-metallic electronic structure) [22] and low effective damping constant ($\alpha$) ~ 0.001 [23], which are beneficial for enhancing the TMR ratio and lowering the magnetization switching current of spin-transfer torque (STT). CFA films prepared using sputtering deposition generally have a disordered $B2$ structure (swapping between $Y$ and $Z$ sites) rather than an ordered $L2_1$ structure owing to the thermodynamic stability of CFA [24]. Nevertheless, the spin polarization calculated for the $L2_1$ structure is conserved even for the $B2$ structure [25]. Importantly, a CFA film with a (001) orientation has a large in-plane lattice spacing ($d_{(200)}/\sqrt{2} = d_{(110)} = 0.203$ nm) compared with other half-metallic Heusler alloys such as $Co_2FeAl_{0.5}Si_{0.5}$ ($d_{(110)} = 0.201$ nm) and $Co_2MnSi$ ($d_{(110)} = 0.198$ nm). Therefore, a nearly perfect CFA/MgO(001) heterostructure is easily achieved by the magnetron sputtering method, and this is favorable for enhancing the coherent tunneling effect [13, 18]. Recently, TMR ratios as high as 360% at RT (785% at 10 K) were demonstrated in epitaxial CFA-based MTJs with a sputter-deposited MgO barrier [17–19]. The large TMR ratio originated from the high spin polarization of the CFA layer and the strong contribution of the coherent tunneling effect through $\Delta_1$ Bloch states in the CFA and the MgO barrier. Moreover, (001)-textured CFA films can be grown on MgO-buffered Si/SiO$_2$ amorphous substrates, and a relatively large TMR ratio of 166% at RT (252% at 48 K) was achieved in a (001)-textured CFA/MgO/CoFe MTJ [20]. Such polycrystalline full-Heusler MTJs on amorphous substrates are desired because of their compatibility with practical industrial applications of full-Heusler spintronic devices, while single-crystal MgO(001) substrates have a limited scope of application.

Furthermore, STT-induced magnetization switching (STT switching), a key technology for writing information in spintronics devices, was realized using MTJs with an epitaxial CFA ultrathin (~1.5 nm) layer as a switching (free) layer [26]. However, a large critical switching current density ($J_{c0}$) of 2.9 $\times$ 10$^7$ A/cm$^2$ was observed owing to the enhancement of $\alpha$, which was ~0.04, of the epitaxial CFA film. In addition, STT switching can be disturbed by the stabilization of intermediate magnetic states possibly because of the presence of in-plane magnetocrystalline anisotropy, which is generally seen in epitaxial magnetic films (e.g., 4-fold anisotropy for cubic (001) films) [27]. Therefore, reducing the undesirable magnetic anisotropy using polycrystalline Heusler alloy films is effective for highly efficient STT switching.

In this work, we systematically studied MTJs with (001)-textured polycrystalline CFA films on Si/SiO$_2$ amorphous substrates. A MgO buffer was introduced in the MTJs for achieving (001) texture with $B$2-ordering structure of CFA layers on the amorphous substrates. The (001)-texture, $B$2 order, and surface morphology of the polycrystalline CFA films and the TMR effect in the entire MTJ stacks were characterized for varying MgO buffer thickness. Also, the MgO-barrier thickness and resistance-area product ($RA$) dependence of the TMR ratios were investigated for the polycrystalline CFA-MTJs. Furthermore, STT switching was examined in low-$RA$ MTJs with a thin polycrystalline CFA film (2.0 nm) as a free layer. $J_{c0}$ of 8.2 $\times$ 10$^6$ A/cm$^2$ was demonstrated by a thermal activation model for switching current; this is far lower than the value for epitaxial CFA-MTJs [26]. The $\alpha$ values for the polycrystalline CFA films, obtained using a waveguide-based ferromagnetic resonance method, were almost constant against the CFA thickness, and a relatively low $\alpha$ of ~0.015 was demonstrated. We attributed the reduction in the $J_{c0}$ of the STT switching to the reduced $\alpha$ of the polycrystalline CFA films.

## II. EXPERIMENT

All multilayer stacks were deposited on thermally oxidized Si/SiO$_2$ amorphous substrates at RT using an ultra-high vacuum magnetron sputtering system with a base pressure lower than $4 \times 10^{-7}$ Pa. MgO layers were deposited from a sintered MgO target by RF sputtering with an RF power density of 2.19 W/cm$^2$ and an Ar pressure of 10 mTorr. CFA layers were deposited from a stoichiometric Co$_{50}$Fe$_{25}$Al$_{25}$ (at.%) alloy target using DC power. The structural properties and surface morphology of CFA films on MgO buffers were characterized by out-of-plane (2$\theta$-$\omega$ scan) X-ray diffraction (XRD) with Cu $K\alpha$ radiation ($\lambda$ = 0.15418 nm) and atomic force microscopy (AFM), respectively. MTJ stacks with the structure of CFA/MgO/Co$_{75}$Fe$_{25}$/IrMn/Ru (unit: nm) were deposited on MgO-buffered thermally oxidized Si/SiO$_2$ amorphous substrates and patterned into junctions with active area of $5 \times 10$ μm$^2$ by conventional lithography methods with Ar ion milling. For STT switching, spin-valve MTJs with the structure of MgO (7.5)/Cr (40)/CFA (2)/MgO (0.6–0.8)/Co$_{75}$Fe$_{25}$ (5)/Ru (0.8)/Co$_{75}$Fe$_{25}$ (5)/IrMn (15)/Ta (5)/Ru (10) (unit: nm) were prepared on the amorphous substrates and nanofabricated into 100-nm-scaled ellipses. The actual areas of the MTJ nano-pillars were obtained according to the ratio of the *RA* to the junction resistance; the *RA* was characterized by current-in-plane tunneling (CIPT) measurement [28] before patterning. The MTJ stacks were post-annealed in a vacuum furnace for 30 minutes under a magnetic field of 5 kOe. The magneto-transport properties were measured using a DC 2- or 4- probe method. The magnetic damping constant $\alpha$ of the polycrystalline CFA film was measured by waveguide-based ferromagnetic resonance (FMR). The films were patterned into rectangular shape elements of 600 × 20 μm² using UV lithography together with Ar ion milling, and then coplanar waveguides made of Au were fabricated on them. The FMR signal obtained as change of the real part of S21 signal, was determined using a network analyzer. An external magnetic field along the longitudinal axis was varied from 0 to 1.9 kOe, while the excitation power was fixed as 0 dBm. All measurements were performed at RT.

## III. RESULTS

### A. Effect of MgO buffer on polycrystalline CFA films and TMR

Before MTJ multilayer films were grown, MgO buffer was deposited on a Si/SiO$_2$ amorphous substrate in order to establish the (001)-texture of polycrystalline CFA films by taking advantage of the unique (001)-texture property of MgO layers on an amorphous substrate. CFA films were subsequently grown on the MgO buffer layer, and the structural properties of the CFA films depending on the MgO buffer thickness were investigated by XRD. The out-of-plane XRD patterns of 30-nm-thick CFA films are shown in Fig. 1(a); the CFA films were annealed at $T_a$ = 400 ℃, and the thicknesses of the MgO buffers ($t_{MgO}$) are 2.5, 5.0, 7.5, and 10.0 nm. In addition to the peaks (denoted by "s") from Si/SiO$_2$ substrates, MgO(002), CFA(002), and (004) peaks were observed along with the absence of other oriented peaks, demonstrating the (001)-texture established in the stacks. Figure 1(b) shows the $t_{MgO}$ dependence of the integrated intensity of CFA(002) peaks for as-deposited CFA films, and CFA films annealed at 400 ℃ and 480 ℃. With increasing $t_{MgO}$, the intensity of the peaks initially increases for all samples owing to the improved (001)-texture of the MgO buffer layer, reaching a maximum at $t_{MgO}$ = 7.5 nm. The reduction in intensity at $t_{MgO}$

= 10 nm may be caused by the degraded surface morphology of the MgO buffer layer. In addition, the XRD intensity increases with increasing annealing temperature, which indicates that the increasing temperature improves the $B2$ order and (001)-texture of the CFA films. In the XRD $2\theta$-$\omega$ scan with the diffraction vector along CFA[111], (111) reflection was not detected, indicating that the CFA films have a $B2$-ordering structure with swapping between Fe and Al atoms while Co atoms occupy the regular sites. The degree of $B2$ ordering was estimated according to the ratio of the integrated intensity of the CFA(002) and (004) peaks. The peaks were fitted by Voigt profiles, and the ratio of their integrated intensities, i.e., the ratio of $I(002)$ to $I(004)$, is shown in Fig. 1(c). The maximum $I(002)/I(004)$ value was obtained at $t_{MgO}$ = 7.5 nm for all of the samples: as-deposited and annealed at 400 ℃ or 480 ℃. This value is comparatively large for CFA films annealed at 480 ℃, indicating the improvement of the $B2$ ordering and the mosaicity of the CFA films due to annealing at high temperature. The degree of $B2$ ordering, $S_{B2}$, can be evaluated using the ratios according to the following equation: [29]

$$S_{B2} = \sqrt{\frac{[I(002)/I(004)]_{exp.}}{[I(002)/I(004)]_{cal.}}}, \qquad (1)$$

where $[I(002)/I(004)]_{exp.}$ is the ratio of the integrated intensity of the (002) peak to that of the (004) peak as determined by experiments, and $[I(002)/I(004)]_{cal.}$ is the ideal ratio of the two peaks. For $t_{MgO}$ = 7.5 nm, the ordering parameter $S_{B2}$ is calculated to be 0.89, 0.95, and 0.98 for CFA films as deposited, annealed at 400 ℃, and 480 ℃, respectively. The results demonstrate a high $B2$ order and an excellent (001)-texture were established in the polycrystalline CFA films on MgO-buffered Si/SiO$_2$ amorphous substrates.

For stacking MTJ multilayers with the polycrystalline CFA films, the $t_{MgO}$ dependence of the surface morphology of the CFA films was investigated. Figure 2 shows the average surface roughness ($Ra$) and peak-to-valley ($P$-$V$) value as a function of $t_{MgO}$ for 30-nm-thick CFA films annealed at 400 ℃ and 480 ℃, respectively. For the samples annealed at 400 ℃, flat surfaces with $Ra$ ~ 0.1 nm and $P$-$V$ ranging from 1.3 to 1.5 nm were observed for all values of $t_{MgO}$. The inset of Fig. 2 shows an example of AFM images of the samples (annealed at 400 ℃ on a 7.5-nm-thick MgO buffer). The results indicate the feasibility of stacking MTJs with a thin MgO barrier. In addition, the samples annealed at 480 ℃ with higher $Ra$ and $P$-$V$ values were observed as well as a large $t_{MgO}$ dependence.

The whole MTJ stacks with the structure of MgO ($t_{MgO}$)/CFA (30)/MgO ($t_{barr}$)/Co$_{75}$Fe$_{25}$ (5)/IrMn (15)/Ru (10) (unit: nm) were then fabricated on Si/SiO$_2$ substrates with varying MgO buffer thickness $t_{MgO}$ (2.5–10.0 nm) and MgO barrier thickness $t_{barr}$ (1.5, 1.8, and 2.0 nm). The stacks were annealed at 370 ℃ in the presence of a magnetic field of 5 kOe. Figure 3 shows TMR ratios as a function of $t_{MgO}$ for the polycrystalline CFA-MTJs measured at RT using CIPT. The 30-nm-thick CFA films were post-annealed at $T_a$ = 400 ℃ and 480 ℃ in order to improve the $B$2 ordering. The TMR ratios obtained in MTJs with $T_a$ = 400 ℃ were higher than those of MTJs with $T_a$ = 480 ℃, which can be attributed to a better CFA/MgO-barrier interface due to the flat CFA surface annealed at 400 ℃, although a higher degree of $B$2 ordering was observed for $T_a$ = 480 ℃, as shown in Figs. 1 and 2. With increasing $t_{MgO}$, the TMR ratio increases and is nearly saturated at $t_{MgO}$ > 5 nm, indicating that high-quality CFA films with $B$2 order and (001)-texture were established with the more than 5-nm-thick MgO buffer layer, which is consistent with the XRD analyses. A slight reduction in the TMR ratio was observed at $t_{MgO} = 10$ nm, which could be caused by the reduction in the degree of (001)-orientation of the MgO buffer. The MTJs with a 1.8-nm-thick MgO barrier exhibit larger TMR ratios than those with 1.5- and 2.0-nm-thick MgO barriers, which could be due to the plastic relaxation of the MgO barrier [30] and/or the oscillatory behavior of the TMR ratio as a function of MgO thickness [18].

## B. MgO barrier thickness and *RA* dependences of TMR

In order to realize STT switching in the polycrystalline CFA-MTJs, it is expected that introduction of a conductive underlayer, reduction in the free-layer thickness, and control of the *RA* of the barrier layer are required. The MgO barrier thickness and *RA* dependences of the TMR ratio were investigated using spin-valve MTJs with the structure of MgO buffer (7.5)/Cr (40)/CFA (30)/MgO ($t_{barr}$: 1.2-2.0)/Co$_{75}$Fe$_{25}$ (5)/IrMn (15)/Ta (5)/Ru (10) (unit: nm) on a Si/SiO$_2$ substrate. A Cr underlayer was selected as the conductive electrode because Cr has a very small lattice mismatch with CFA (~0.6%) and can further facilitate the ordering structure of full-Heusler alloys [13]. The Cr layers for the samples as deposited, annealed at 400 ℃, and 600 ℃ were prepared on the 7.5-nm-thick MgO buffer for the MTJ stacks. The entire stacks were annealed at 370 ℃ in the presence of a magnetic field of 5000 Oe, and then their TMR ratios and *RA* values were characterized using CIPT measurement.

Figure 4(a) shows the dependence of the TMR ratios on the nominal thickness of the MgO barrier ($t_{barr}$) for the MTJs with different Cr annealing conditions. For the samples that were as-deposited and annealed at 400 ℃, the TMR ratio increases with $t_{barr}$, and TMR ratios greater than 100% were achieved for the whole range of $t_{barr}$. This means that the (001)-texture and *B*2 ordering of CFA films can be maintained on the MgO/Cr buffer layers. We obtained the largest TMR ratio of 175% for the 400 °C annealed sample with $t_{\text{barr}}$ = 1.95 nm; this TMR ratio is higher than 166%, which was observed in the MTJ without the Cr buffer, which indicates that the Cr buffer with optimal conditions promotes CFA(001) growth and improves the effective tunneling spin polarization. On the other hand, the samples annealed at a high temperature (600 °C) exhibited smaller TMR ratios (80–120%). This was attributed to the rough surface (*Ra* = 0.4 nm and *P-V* = 3.2 nm) of the CFA film on the Cr layer annealed at 600 °C, which can lead to a declined crystalline orientation of the MgO(001) barrier. Furthermore, oscillation behavior of the TMR ratios as a function of $t_{\text{barr}}$ was observed for all of the structures. The TMR oscillation behavior is typically observed in epitaxial MTJs such as Fe/MgO/Fe [31], $Co_2MnSi/MgO/Co_2MnSi$ [32], $Co_2Cr_{0.6}Fe_{0.4}Al/MgO/Co_2Cr_{0.6}Fe_{0.4}Al$ [33], and CFA/MgO/CoFe [18] MTJs, while it is absent in polycrystalline MTJs such as CoFeB/MgO/CoFeB [34] MTJs. More remarkable oscillation amplitude in epitaxial full-Heusler alloy-based MTJs than that of epitaxial Fe/MgO/Fe MTJs was observed, which may be related to the electronic structures of full-Heusler alloy electrodes and the full-Heusler/MgO interface; however, the origin has not been understood yet. The unexpected oscillation behavior in the polycrystalline CFA-MTJs may be also attributed to the unique electronic structure of CFA and the interface. In addition, the flat buffer layer with a good crystallinity enabled us to achieve a well-defined layer-by-layer growth for the CFA layer and the MgO barrier, which may be advantageous for observing the oscillatory behavior. The oscillation period was approximately 0.2 nm in nominal thickness, which seems to be shorter than that for the epitaxial CFA/MgO/CoFe (0.32 nm, short-period) [18]. Further investigation is needed to clarify the origin of the behavior.

The *RA* as a function of the MgO barrier thickness is plotted in Fig. 4(b). We observed a typical behavior of an exponential increase with increasing $t_{barr}$. According to the Wenzel-Kramer-Brillouin (WKB) approximation, the relationship between *RA* and $t_{barr}$ can be expressed as follows:

$$RA(t_{barr}) \propto exp(\frac{4\pi\sqrt{2m\phi}}{h} t_{barr}), \qquad (2)$$

where *h*, *m*, and $\phi$ are Planck's constant, the effective electron mass assumed as a free electron mass ($9.11 \times 10^{-31}$ kg) here, and the barrier height energy of the tunnel barrier, respectively [31]. A similar barrier height of 0.7 eV was obtained for all three samples by the fitting of *RA*-$t_{barr}$ curves. This value is greater than the reported values for Fe/MgO/Fe grown using molecular-beam-epitaxy (MBE) (0.39 eV) [31], sputtered CoFeB/MgO/CoFeB (0.29–0.39 eV) [35–37], and CoFeB/MgO (electron-beam evaporated)/CoFeB (0.48 eV) [38] MTJs. The reasons may be due to the different densities of oxygen vacancy defects in the MgO barriers and/or the deviation of the actual MgO thickness from the nominal one.

Figure 5 shows TMR ratios in a low-*RA* regime for the polycrystalline CFA-MTJs with 2-nm-thick CFA film as a free layer. The MTJs were annealed at 225 °C for 30 minutes in order to reduce the influence of Cr layer to CFA. A TMR ratio of 40–60% was achieved with an *RA* of 7–20 Ω μm$^2$ (nominal MgO thickness: 0.6–0.8 nm) for the 100-nm-scaled elliptical MTJs with the thin CFA layer, which is favorable for achieving STT switching in the MTJ stacks with a thin CFA free layer and the low *RA* value. In addition, the TMR ratio of the polycrystalline MTJs with 2-nm-thick CFA film is comparable to that with a thick (30 nm) CFA film at a low *RA* value, as shown in Fig. 5, which indicates that the (001)-texture and *B*2 ordering can be maintained in the thin 2-nm-thick CFA films.

### C. STT-induced magnetization switching

STT-induced magnetization switching was performed in spin-valve MTJs with the structure of Si/SiO$_2$-substrate/MgO (7.5)/Cr (40)/CFA (2)/MgO (0.6–0.8)/Co$_{75}$Fe$_{25}$ (5)/Ru (0.8)/Co$_{75}$Fe$_{25}$ (5)/IrMn (15)/Ta (5)/Ru (10) (unit: nm). A schematic of the structure of a polycrystalline CFA-MTJ nanopillar is shown in Fig. 6(a). The synthetic antiferromagnetic coupling exchange bias of CoFe/Ru/CoFe/IrMn was employed to reduce the offset magnetic field of hysteresis loops. Figure 6(b) indicates the tunneling resistance of an MTJ nanopillar as a function of the applied magnetic field (*H*) measured with a DC bias voltage of 1 mV. A TMR ratio of 43% was observed in the MTJ with a thin CFA free layer (2.0 nm) and MgO barrier (0.75 nm). Sharp switching between parallel (P) and antiparallel (AP) magnetic configurations was observed. The *RA* of the MTJ was determined using CIPT measurements to be 13 Ω·μm$^2$, and the active area of the MTJ nanopillar was calculated to be 1.24 × 10$^{-2}$ μm$^2$. The hysteresis offset field ($H_{\text{offset}}$) and the coercivity field ($H_c$) were determined using the *R-H* loops to be −11 and 26 Oe, respectively. Figure 6(c) shows the representative resistance-current (*R-I*) loops of the CFA-MTJ nanopillar measured by a DC current with a sweep rate of 1.2 × 10$^{-4}$ A/s at different magnetic fields of 0, −11 and −20

Oe, respectively. The positive current indicates that electrons flow from the bottom electrode to the top electrode. Magnetic switching between P (low-resistance) and AP (high-resistance) states was achieved owing to the current. When current was applied in the negative (positive) direction, the P (AP) state can be obtained from AP (P) state, corresponding to magnetization reversal of the CFA free layer. Also, the critical switching currents ($I_c$) in both directions shift in the negative direction with the decrease of magnetic field from 0 to −20 Oe. These results indicate typical behaviors of STT-induced magnetization switching.

Since the STT switching by the DC current is a thermally activated process [39–41], we use a thermal activation model for switching currents deduced from *R-I* loops to evaluate the intrinsic critical switching current density ($J_{c0}$) and thermal stability factor $\Delta_0$ (= $K_u V/k_B T$) for the MTJ, where $K_u$ is the uniaxial magnetic anisotropy, $V$ is the volume of the free layer, $k_B$ is the Boltzmann constant, and $T$ is the absolute temperature. In the thermal activation model, the sweep current $I(t)$ is assumed to increase linearly with time $t$, i.e., $I(t) = vt$, and the cumulative probability distribution function $P(t)$ of the switching current in $H_{\text{offset}}$ can be expressed as

$$P(t) = 1 - \exp\left(-\frac{f_0 I_{c0}}{v \Delta_0} \left\{ \exp\left[-\Delta_0 \left(1 - \frac{vt}{I_{c0}}\right)\right] - \exp[-\Delta_0] \right\}\right), \quad (3)$$

where $f_0$ is the effective attempt frequency (=$10^9$ Hz), $I_{c0}$ is the intrinsic switching current, and $v$ is a constant sweep rate of the sweep current in the measurement of *R-I* loops [41]. The distribution of the critical switching current $I_c$ was obtained by repeating the measurement of the *R-I* loops for 300 times. Figure 6(d) shows typical *R-I* loops at $H_{\text{offset}}$ =

−11 Oe for the polycrystalline CFA-MTJ nanopillar. The mean critical currents in the positive ($I_{c,P \to AP}$) and negative ($I_{c,AP \to P}$) directions are determined to be 530 and 400 μA, corresponding to the critical current density of $4.3 \times 10^6$ ($J_{c,P \to AP}$) and $3.2 \times 10^6$ A/cm$^2$ ($J_{c,AP \to P}$), respectively. Figures 6(e) and (f) show the switching probability for $I_{c,P \to AP}$ and $I_{c, AP \to P}$ as a function of the sweep current. Using the constant sweep rate $v = 1.2 \times 10^{-4}$ A/s in the measurement of R-I loops, the switching probability was fitted using Eq. (3), as shown by the solid lines. As a result, the intrinsic current density of $J_{c0,P \to AP}$ ($J_{c0,AP \to P}$) = 9.1 $\times 10^6$ A/cm$^2$ ($7.3 \times 10^6$ A/cm$^2$) and thermal stability of $\Delta_{0,P \to AP}$ ($\Delta_{0,AP \to P}$) = 30.0 (28.4) were achieved for the polycrystalline CFA-MTJ nanopillar.

### D. Gilbert damping of the polycrystalline CFA film

The Gilbert damping parameter, $\alpha$, is a critical parameter for determining the critical current density of STT switching. In order to examine $\alpha$ for the polycrystalline CFA film, waveguide-based FMR was performed in a single ferromagnetic layered sample, consisting of SiO$_2$-substrate//MgO(7.5 nm)/Cr(40 nm)/CFA(2-18 nm) structure. Typical FMR spectra with varied external magnetic fields ($H_{ext}$) for a sample of 18-nm-thick CFA are shown in the inset of Fig. 7(a). A clear shift in the resonant frequency can be seen as $H_{ext}$ increases from 500 to 1500 Oe. The peak intensity is relatively small at low magnetic field, possibly because of the anisotropy distribution inside the film.

Figures 7(a) and (b) show the $H_{ext}$ dependence of the resonant frequency ($f_0$), demagnetization field ($H_d$), and magnetic damping parameter ($\alpha_H$), respectively, estimated by fitting each spectrum using an analytical solution [42]. Here, we assume the gyromagnetic ratio ($\gamma$) as $2\pi \times 0.00297$ GHz/Oe and neglect the in-plane magnetic anisotropic field. Both $H_d$ and $\alpha_H$ exhibit a weak dependence on $H_{ext}$ possibly due to the anisotropy distribution, which becomes saturated at a high magnetic field range. The $f_0$ was fitted using the simplified Kittel formula:

$$f_0 = \frac{\gamma}{2\pi}\sqrt{H_{ext}(H_{ext}+H_d)} \qquad (4)$$

Then, we obtained $H_d = 12154 \pm 17.9$ Oe, which agrees well with the values obtained in the individual resonant spectrum for $H_{ext} > 300$ Oe. The magnetic field dependence of $\alpha$ can be excluded by the fitting equation,

$$\alpha_H = \alpha + k\exp(-H_{ext}/H_0), \qquad (5)$$

where $k$ and $H_0$ are fitting parameters, and $\alpha$ is estimated to be $0.0148 \pm 0.0003$. Figure 7(c) summarizes CFA thickness dependence of the saturation magnetization, $M_s$, and Gilbert damping constant, $\alpha$, of polycrystalline CFA films. The weak thickness dependence of $M_s$ and $\alpha$ indicates that the CFA film guarantees a good quality even in a thin thickness regime at around 2 nm.

## IV. DISCUSSION

Using polycrystalline CFA full-Heusler thin films, MTJs with the structure of CFA/MgO/Co$_{75}$Fe$_{25}$ were successfully fabricated on a SiO$_2$ amorphous substrate. The effects of the MgO buffer on the structural properties of the polycrystalline CFA films and the TMR in the MTJs were investigated. Optimized (001)-texture, $B2$ order, and surface morphology of the CFA films was demonstrated on a 7.5-nm-thick MgO buffer, which contributes a large TMR ratio in the whole polycrystalline MTJ stacks. In order to achieve STT switching, Cr underlayers were utilized as a conductive electrode on the optimized MgO buffer. The Cr layer was known to have very small lattice mismatch (~0.6%) with CFA and facilitate the ordering structure of full-Heusler alloys. A TMR ratio of 175% was achieved in the polycrystalline CFA-MTJs on MgO/Cr-buffered Si/SiO$_2$ substrate. The buffer layer dependence of the structural properties of polycrystalline CFA films and the TMR ratios in entire MTJ stacks are significant for practical spintronic applications of full-Heusler alloy materials. A proper buffer layer with minimal diffusion, and enhanced (001)-texture and ordering parameter of polycrystalline full-Heusler alloys is required for further increasing the TMR ratios of the MTJs.

STT switching was performed in the CFA-MTJs with polycrystalline CFA as a free layer. The average intrinsic current density $J_{c0} = (J_{c0,P \to AP} + J_{c0,AP \to P})/2 = 8.2 \times 10^6$ A/cm$^2$ for the polycrystalline CFA-MTJ is generally comparable to that reported for CoFeB-MTJs with in-plane magnetization [43, 44]; however, it is much lower than the value of $2.9 \times 10^7$ A/cm$^2$ for epitaxial CFA-MTJs [26]. Based on Slonczewski's model of STT switching [45, 46], the simplified $J_{c0}$, ignoring external magnetic fields, is given by,

$$J_{c0} = \left(\frac{2e}{\hbar}\right)\frac{\alpha M_s t}{\eta} H_{eff}, \qquad (6)$$

where $e$ is the electron charge, $\hbar$ is reduced Planck's constant, $M_s$ is the saturation magnetization, $t$ is the thickness of the free layer, $\eta$ is the spin-transfer efficiency, and $H_{eff}$ the effective field acting on the free layer, including the magnetocrystalline anisotropy field, demagnetization field, stray field. The polycrystalline CFA free layer ($t$ = 2.0 nm) is thicker than the epitaxial one ($t$ = 1.5 nm). A similar $M_s$ (~1000 emu/cm$^3$) at RT was observed for both polycrystalline and epitaxial CFA films. The $\eta$ of the polycrystalline CFA-MTJs should be smaller than that of the epitaxial CFA-MTJs because the TMR ratio of the polycrystalline CFA-MTJs (43%) is lower than that of the epitaxial CFA-MTJs (60%) [26]. Accordingly, the low $J_{c0}$ in the polycrystalline CFA-MTJs can be mainly attributed to the small $\alpha$ of the polycrystalline CFA free layer (~0.015) compared with that of the epitaxial CFA films (~0.04) [26]. For the polycrystalline CFA-MTJs, annealing was performed with a low temperature of 225 ℃ and a short time of 30 minutes, whereas the epitaxial CFA-MTJs were annealed at 360 ℃ for 1 hour; this could be a factor in the reduction in $\alpha$ [47]. As a result, a lower $J_{c0}$ was obtained in the polycrystalline CFA-MTJs than that in the epitaxial ones. However, the $\alpha$ value is still greater than that reported for a 50-nm-thick CFA film on a MgO layer annealed at 600 °C (~0.001) [23]; this may be attributed to the inter-diffusion of Cr atoms into the CFA layer and the residual magnetic moments on the Cr surface. Consequently, a proper buffer material is strongly required for CFA full-Heusler MTJs in order to reduce the α of the free layer and thus the $J_{c0}$ of STT switching. Another factor for reducing $J_{c0}$ is the magnetic anisotropy of the free layer. In perpendicular anisotropy CoFeB/MgO/CoFeB tunnel junctions, a low current density for STT switching was demonstrated owing to the perpendicular magnetic anisotropy (PMA) [48-53]. To evaluate the contribution of interface PMA, magnetization measurements at in-plane and out-of-plane of the polycrystalline CFA film was performed. An effective anisotropy energy density ($K_{eff}$) of −5 × 10$^6$ erg/cm$^3$ was obtained for a 2-nm-thick CFA/MgO structure where the negative $K_{eff}$ indicates the CFA layer is in-plane magnetized. In general, $K_{eff}$ can be simply expressed by the equation of $K_{eff} = K_v + K_i/t$, where $K_v$ is the volume anisotropy energy density which can be treated as demagnetization energy density ($2\pi M_s^2$) for simplicity, $K_i$ is the interface anisotropy energy density, and $t$ is the thickness of the CFA layer. As a result, the value of $K_i$ can be calculated to be 0.25 erg/cm$^2$, indicating that an interface PMA is induced at the CFA/MgO interface. In the CFA/MgO-MTJs, the interface PMA can cause a reduced effective anisotropy, and may also play a role for the reduction in $J_{c0}$. For further decreasing $J_{c0}$, a chosen buffer material, out-of-plane magnetization of the CFA free layer [54–56], and/or advanced fabricating techniques for high-quality CFA-MTJs are required.

## V. CONCLUSION

In conclusion, TMR ratios and STT-induced magnetization switching were studied in (001)-textured polycrystalline CFA full-Heusler based MTJs on Si/SiO$_2$ amorphous substrates. CFA films with a good (001)-texture and high $B$2 order were achieved on MgO-buffered Si/SiO$_2$ amorphous substrates. The MgO barrier thickness and $RA$ dependences of the TMR ratio in the polycrystalline CFA-MTJs were also studied. Moreover, STT switching was achieved in the MTJs with a thin polycrystalline CFA film (2.0 nm) as a free layer. The $J_{c0}$ of $8.2 \times 10^6$ A/cm$^2$ was demonstrated for the polycrystalline CFA-MTJs with in-plane magnetization using a thermal activation model for a cumulative switching probability distribution with a sweep current. The $J_{c0}$ is much lower than that reported for epitaxial CFA-MTJs, which is mainly attributed to the reduced $\alpha$ of the polycrystalline CFA free layer.

ACKNOWLEDGMENTS

This work was partly supported by the Japan Science and Technology Agency, CREST, and JSPS KAKENHI Grant Number 23246006.

**Figures and captions:**

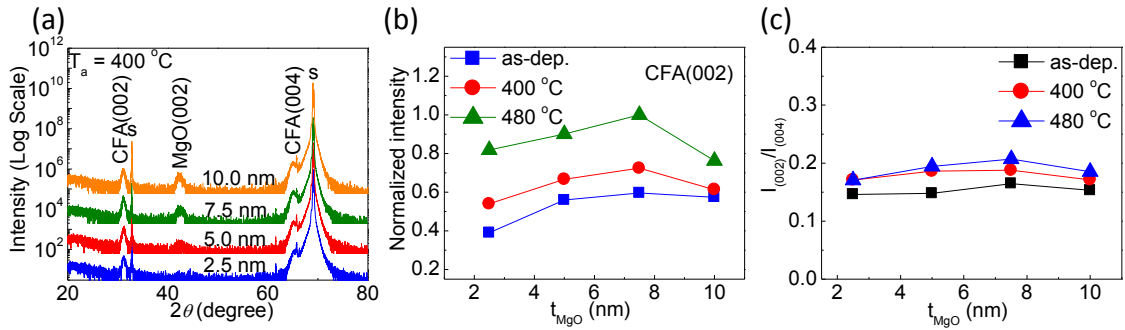

FIG. 1. (Color online) (a) Out-of-plane ($2\theta$-$\omega$ scan) XRD patterns for polycrystalline CFA full-Heusler alloy films on MgO-buffered $Si/SiO_2$ amorphous substrates with varied MgO buffer thickness, $t_{MgO}$: 2.5, 5.0, 7.5, and 10.0 nm. (b) Normalized integrated intensity of CFA(002) peak, and (c) Ratios of (002) to (004) peaks as a function of $t_{MgO}$ for as-deposited (as-dep.) samples and those annealed at 400 ℃ and 480 ℃.

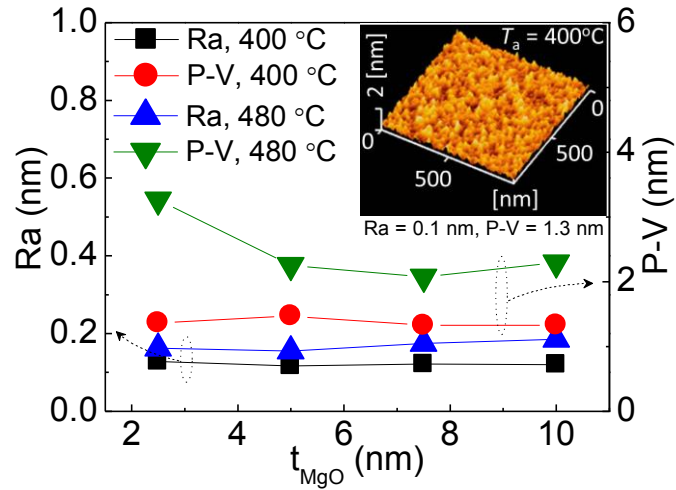

FIG. 2. (Color online) Surface morphology of polycrystalline CFA full-Heusler alloy films on MgO-buffered Si/SiO$_2$ amorphous substrates with respect to the thickness of the MgO buffer. The CFA films were 30-nm-thick and were deposited at RT and post-annealed at 400 °C and 480 °C, respectively. Inset: AFM image of the surface of the CFA film annealed at 400 °C.

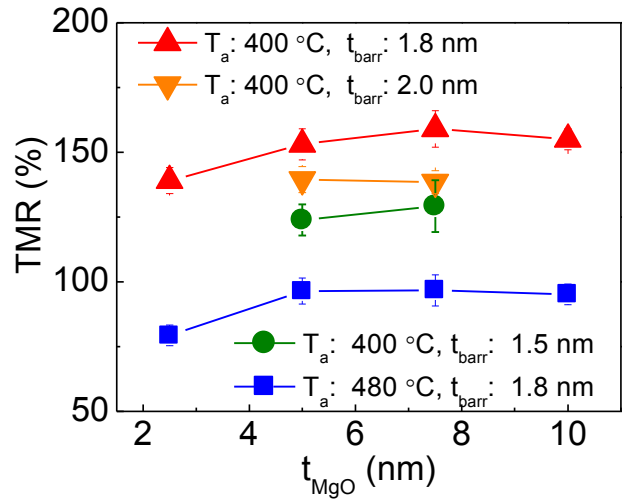

FIG. 3. (Color online) TMR ratios as a function of MgO buffer thickness $t_{MgO}$ for polycrystalline CFA/MgO/CoFe MTJs with different thicknesses of the MgO barrier, $t_{barr}$: 1.5 nm, 1.8 nm, and 2.0 nm. The 30-nm-thick CFA bottom electrodes were directly deposited on the MgO buffer layer and post-annealed at 400 ℃ and 480 ℃ after deposition at RT. The MTJ stacks were annealed at 370 ℃ before the CIPT measurement.

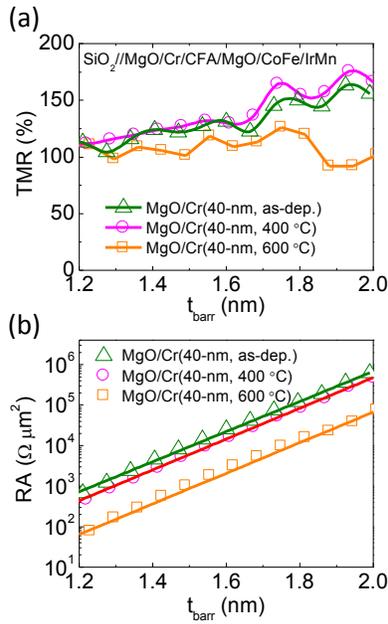

FIG. 4. (Color online) The thickness of MgO barrier, $t_{barr}$, dependence of (a) TMR ratios and (b) *RA* at RT for polycrystalline CFA/MgO/CoFe MTJs with as-deposited, 400 ℃ and 600 ℃ annealed MgO(7.5 nm)/Cr(40 nm) buffer layers on $SiO_2$ amorphous substrates, characterized using CIPT measurement.

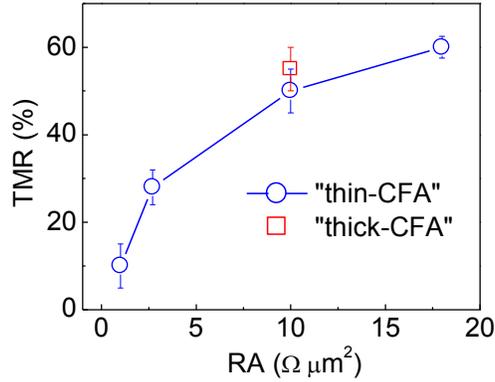

FIG. 5. (Color online) *RA* dependence of TMR ratios for polycrystalline CFA-MTJs with a 2-nm-thick CFA layer ("thin-CFA") as a bottom electrode. The squared symbol indicates the TMR ratio for a polycrystalline CFA-MTJ with a 30-nm-thick CFA layer ("thick-CFA").

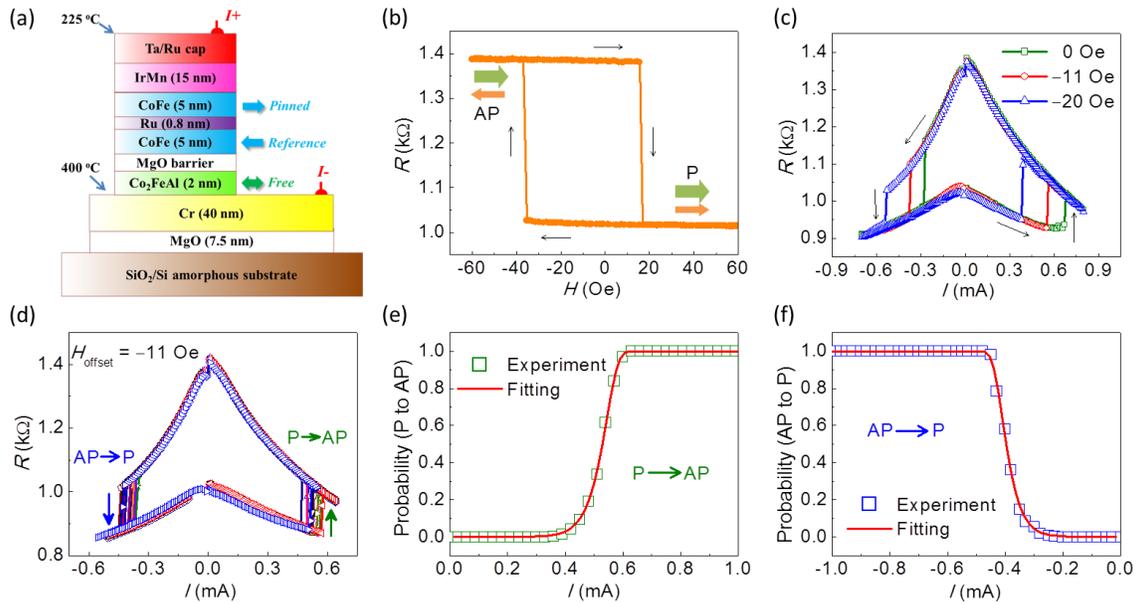

FIG. 6. (a) Schematic illustration of the structure of polycrystalline CFA-MTJ nanopillar. (b) *R-H* loops for a polycrystalline CFA-MTJ nanopillar. Wide arrows show the magnetic configurations of bottom (free) and top (reference) electrodes of the MTJ, and narrow arrows indicate sweep direction of the applied magnetic field. (c) *R-I* loops for the MTJ at magnetic fields of 0, −11 and −20 Oe, respectively. Arrows indicate sweep direction of the applied current. (d) Representative *R-I* loops of the CFA-MTJ at applied magnetic field of −11 Oe. (e) and (f) Switching probabilities for $I_{c,P\to AP}$ and $I_{c,AP\to P}$ obtained by repeating *R-I* measurements for 300 times. Solid lines are fitting curves given by Eq. (3). All measurements were performed at RT.

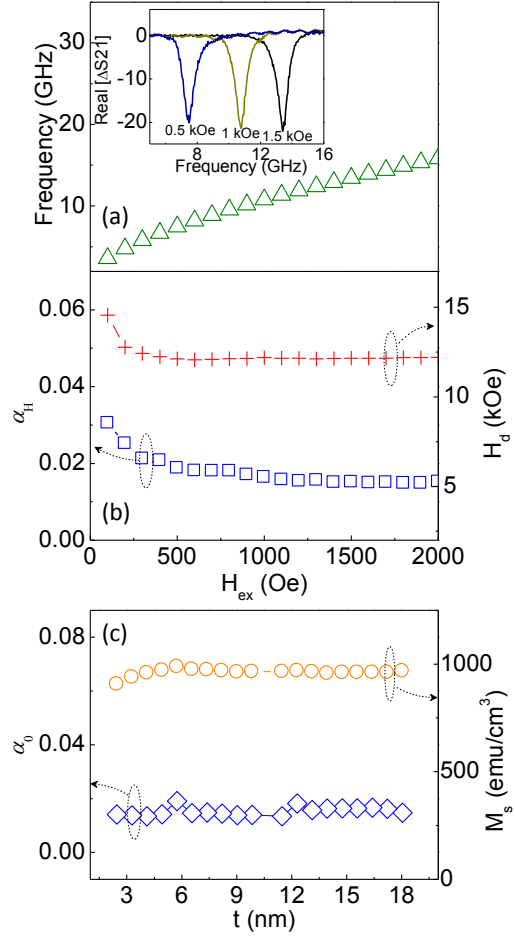

FIG. 7. The $H_{ext}$ dependence of (a) resonant frequency $f_0$, (b) demagnetization field $H_d$, and magnetic damping parameter $\alpha_H$ estimated by fitting the FMR spectrum at each magnetic field for a sample of a 18-nm-thick polycrystalline CFA film. Inset of (a) is typical FMR spectra at $H_{ext}$ of 500, 1000, and 1500 Oe. (c) CFA thickness *t* dependence of saturation magnetization $M_s$ and damping constant $\alpha_0$ of the polycrystalline CFA film.